\documentclass[]{article}

\usepackage{paralist}

\usepackage{calrsfs}

\usepackage{oldgerm}

\usepackage{amssymb,amsbsy}
\usepackage{epic,eepic}
\usepackage{texdraw}
\usepackage{dsfont}
\usepackage{latexsym}

\usepackage{ntheorem}

\usepackage{accents}
\DeclareMathAccent{\wtilde}{\mathord}{largesymbols}{"65}
\DeclareMathAccent{\what}{\mathord}{largesymbols}{"62}

\makeatletter
\def\wb{\accentset{{\cc@style\underline{\mskip10mu}}}}
\makeatother

\newcommand{\be}{\begin{equation}}
\newcommand{\ee}{\end{equation}}
\newcommand{\bdm}{\begin{displaymath}}
\newcommand{\edm}{\end{displaymath}}

\theoremstyle{break}

\newtheorem{proposition}{Proposition}[section]

\theorembodyfont{\upshape}

\newtheorem{remark}{Remark}[section]
\newtheorem*{proof}{Proof}

\begin{document}

\title{Integrability and Symmetries of Difference Equations: the Adler--Bobenko--Suris Case}
\author{P. Xenitidis\\
Department of Mathematics, University of Patras, 265 00 Patras, Greece\\}

\maketitle

E-mail : xeniti@math.upatras.gr



\begin{abstract}
We consider the partial difference equations of the Adler-Bobenko-Suris classification, which are characterized as multidimensionally consistent. The latter property leads naturally to the construction of auto-B{\"a}cklund transformations and Lax pairs for all the equations in this class. Their symmetry analysis is presented and infinite hierarchies of generalized symmetries are explicitly constructed.
\end{abstract}

\section{Introduction}

It is well known that, integrable differential equations, like the Korteweg-de Vries (KdV), the sine-Gordon and the nonlinear Schr{\"o}dinger, have many properties in common. They can be written as the compatibility condition of a Lax pair, which plays a crucial role in solving the initial value problem by the inverse scattering transform. They admit auto-B{\"a}cklund transformations, which allow us to construct new solutions from known ones \cite{Xenitidis:ablclar}.

Other properties arise from the symmetries of these equations. Specifically, they admit infinite hierarchies of generalized symmetries and, consequently, infinite conservation laws \cite{Xenitidis:Olver}. Also, they reduce to Painlev{\'e} equations and result from reductions of the Yang-Mills equations \cite{Xenitidis:Mason}. All the above mentioned characteristics may be considered as criteria establishing the integrability of a differential equation.

Analogous characteristics seem to be in common among integrable difference equations defined on an elementary square of the lattice. In the most well known cases, these equations are characterized by their ``multidimensional consistency''. For a two dimensional equation, this means that, the equation may be imposed in a consistent way on a three dimensional lattice, and, consequently, on a multidimensional one. This property incorporates some of the above mentioned integrability aspects. Specifically, the consistency property provides the means to derive algorithmically B{\"a}cklund transformation and Lax pair of the difference equation under consideration \cite{Xenitidis:BobSuris,Xenitidis:Nij1}.

Adler, Bobenko and Suris (ABS) used multidimensional consistency as the key property characterizing integrable difference equations to classify integrable scalar difference equations \cite{Xenitidis:ABS,Xenitidis:ABS1}. The equations emerged from the classification \cite{Xenitidis:ABS} split into the {\bf H} and {\bf Q} list and comprise, apart from known ones, a number of new cases. The subsequent study of the resulting equations has led to construction of exact solutions \cite{Xenitidis:AHN1,Xenitidis:AHN2}, B{\"a}cklund transformations \cite{Xenitidis:Atkinson}, symmetries \cite{Xenitidis:RHsym,Xenitidis:TTX} and conservation laws \cite{Xenitidis:RHcl}.

The ABS equations will be used as the illustrative example to investigate the similarities between discrete and continuous integrable equations, as described above. Specifically, we will show how the auto-B{\"a}cklund is inherited by the multidimensional consistency to these equations. Using the well known relation between such transformations and Lax pairs, \cite{Xenitidis:Crampin}, we will derive Lax pairs for the ABS equations.

The second direction of our investigation will be the symmetries of the ABS equations. In particular, we will prove that they admit infinite hierarchies of generalized symmetries, which are constructed inductively using linear differential operators. The latter can be regarded as recursion operators for the higher members of the ABS class, i.e. equations {\bf Q3} and {\bf Q4}.

The paper is organized as follows. In Section 2 we introduce the notation and present the background material on symmetries of difference equations. The next section is devoted to a short description of two classes of difference equations, one of which contains the ABS equations.

Section 4 deals with the integrability aspects of the equations under consideration, and, in particular, with their auto-B{\"a}cklund transformations and Lax pairs. Section 5 describes the symmetries of the ABS equations and how they can be used effectively to construct solutions, as well as infinite hierarchies of generalized symmetries.

\section{Notation and preliminaries on symmetries of difference equations}

We first introduce the notation that it will be used in what follows. Also, we present those definitions on symmetries of difference equations that will be used in the next sections. For details on the subject, we refer the reader to the clear and extended review by Levi and Winternitz \cite{Xenitidis:tp:Levi1a}. 

A partial difference equation is a functional relation among the values of a function $u : {\mathds{Z}} \times {\mathds{Z}} \rightarrow {\mathds{C}}$ at different points of the lattice, which may involve the independent variables $n$, $m$ and the lattice spacings $\alpha$, $\beta$, as well, i.e. a relation of the form
\begin{equation} {\cal E}(n,m,u_{(0,0)},u_{(1,0)},u_{(0,1)},\ldots;\alpha,\beta)\,=\,0\,. \label{Xenitidis:gendisceq} \end{equation}
In this relation, $u_{(i,j)}$ is the value of the function $u$ at the lattice point $(n+i,m+j)$, e.g.
$$u_{(0,0)}\,=\,u(n,m)\,,\quad u_{(1,0)}\,=\,u(n+1,m)\,,\quad u_{(0,1)}\,=\,u(n,m+1)\,,$$
and this is the notation that we will adopt for the values of the function $u$ from now on. 

The analysis of such equations is facilitated by the introduction of two translation operators acting on functions on ${\mathds{Z}}^2$, defined by 
$$\left( \mathcal{S}_n^{(k)} u \right)_{(0,0)} = u_{(k,0)}\,,\quad \left( \mathcal{S}_m^{(k)} u\right)_{(0,0)} = u_{(0,k)}\,,\quad {\mbox{where}}\,\, k \in \mathds{Z} \,,$$
respectively.

Let $\mbox{G}$ be a connected one-parameter group of transformations acting on the dependent variable $u_{(0,0)}$ of the lattice equation (\ref{Xenitidis:gendisceq}) as follows
$${\mbox{G}}\,:\,u_{(0,0)}\,\longrightarrow\,{\tilde u}_{(0,0)}\,=\,\Phi(n,m,u_{(0,0)};\varepsilon)\,,\quad \varepsilon \,\,\in\,\,{\mathds{R}}\,. $$
Then, the prolongation of the group action of $\rm G$ on the shifted values of $u$ is defined by
\begin{equation}
{\rm G}^{(k)}\,:\,(u_{(i,j)}) \,\longrightarrow\,\left({\tilde u}_{(i,j)}\,=\,\Phi(n+i,m+j,u_{(i,j)};\varepsilon)\right)\,.
\label{Xenitidis:eq:gract} \end{equation}

The transformation group ${\rm G}$ is a {\emph{Lie point symmetry}} of the lattice equation (\ref{Xenitidis:gendisceq}) if it transforms any solution of (\ref{Xenitidis:gendisceq}) to another solution of the same equation. Equivalently, ${\rm G}$ is a symmetry of equation (\ref{Xenitidis:gendisceq}),
if the latter is not affected by transformation (\ref{Xenitidis:eq:gract}), i.e.
$${\cal E}(n,m,\tilde{u}_{(0,0)},\tilde{u}_{(1,0)},\tilde{u}_{(0,1)},\ldots;\alpha,\beta)\,=\,0\,. $$

In essence, the action of the transformation group $\rm G$ is expressed by its {\emph{infinitesimal generator}}, i.e. the vector field
$${\mathbf{x}}\,=\,R(n,m,u_{(0,0)})\,\partial_{u_{(0,0)}}\,,$$
where $R(n,m,u_{(0,0)})$ is defined by
$$R(n,m,u_{(0,0)})\,=\, \left . \frac{{\rm{d}} \phantom{\varepsilon}}{{\rm{d}} \varepsilon} \Phi(n,m,u_{(0,0)};\varepsilon) \right| _{\varepsilon = 0}\,,$$
and is referred to as {\emph{the characteristic}}.

Using the latter, one defines the $k$-th order forward prolongation of $\bf x$, namely the vector field
$${\mathbf{x}}^{(k)}\,=\,\sum_{i=0}^{k}\sum_{j=0}^{k-i} \left({\mathcal{S}}_n^{(i)}\circ {\mathcal{S}}_m^{(j)} R\right)(n,m,u_{(0,0)}) \,\partial_{u_{(i,j)}}\,.$$

This allows one to give an infinitesimal form of the criterion for ${\rm G}$ to be a symmetry  of equation (\ref{Xenitidis:gendisceq}). It consists of the condition
\begin{equation}
 {\mathbf{x}}^{(k)} \left({\mathcal{E}}\left(n,m,u_{(0,0)},u_{(1,0)},u_{(0,1)},\ldots;\alpha,\beta\right) \right) \,=\,0\,.
\label{Xenitidis:eq:infcr} \end{equation}
which should hold for every solution of equation (\ref{Xenitidis:gendisceq}) and, therefore, the latter should be taken into account, when condition (\ref{Xenitidis:eq:infcr}) is tested out explicitly.

Equation (\ref{Xenitidis:eq:infcr}) delivers the most general infinitesimal Lie point symmetry of equation (\ref{Xenitidis:gendisceq}). The solutions of (\ref{Xenitidis:eq:infcr}) determine the Lie algebra $\frak{g}$ of the corresponding symmetry group ${\rm G}$ and the latter can be constructed by exponentiation:
$$\Phi(n,m,u_{(0,0)};\varepsilon)\,=\,\exp(\varepsilon {\mathbf{x}}) \,u_{(0,0)}\,.$$

On the other hand, one may consider a group of transformations $\Gamma$ acting, not only on the dependent variable $u$, but on the lattice parameters $\alpha$, $\beta$ as well. This leads to the notion of {\emph{the extended Lie point symmetry}}. The infinitesimal generator of the group action of $\Gamma$ is a vector field of the form
$${\bf v}\,=\,R(n,m,u_{(0,0)})\,\partial_{u_{(0,0)}}\,+\,\xi(n,m,\alpha,\beta)\,\partial_\alpha\,+\,\zeta(n,m,\alpha,\beta)\,\partial_\beta\,, $$
and the infinitesimal criterion for a connected group $\Gamma$ to be an extended Lie point symmetry of equation (\ref{Xenitidis:gendisceq}) is
$${\mathbf{v}}^{(k)} \left({\mathcal{E}}\left(n,m,u_{(0,0)},u_{(1,0)},u_{(0,1)},\ldots;\alpha,\beta\right) \right)\,=\,0\,,$$
where
\begin{eqnarray*}{\mathbf{v}}^{(k)}&=& \sum_{i=0}^{k}\sum_{j=0}^{k-i} \left({\mathcal{S}}_n^{(i)}\circ {\mathcal{S}}_m^{(j)} R\right)(n,m,u_{(0,0)}) \,\partial_{u_{(i,j)}} \\
& & + \,\xi(n,m,\alpha,\beta)\,\partial_\alpha\,+\,\zeta(n,m,\alpha,\beta)\,\partial_\beta\end{eqnarray*}
is the $k$-th forward prolongation of the symmetry generator $\bf v$.

By extending the geometric transformations to the more general ones, which depend, not only on $n$, $m$ and $u_{(0,0)}$, but also on the shifted values of $u$, we arrive naturally at the notion of the {\emph{generalized symmetry}}. For example, a five point generalized symmetry may be given by the vector field
$${\mathbf{v}}\,=\,R(n,m,u_{(0,0)},u_{(1,0)},u_{(0,1)},u_{(-1,0)},u_{(0,-1)}) \partial_{u_{(0,0)}}\,,$$
while an extended five point generalized symmetry is given by the vector field
\begin{eqnarray*}{\mathbf{v}}&=& R(n,m,u_{(0,0)},u_{(1,0)},u_{(0,1)},u_{(-1,0)},u_{(0,-1)}) \partial_{u_{(0,0)}} \\ 
&& +\,\xi(n,m,\alpha,\beta)\,\partial_\alpha\,+\,\zeta(n,m,\alpha,\beta)\,\partial_\beta\,,\end{eqnarray*}
respectively.

\section{Classes of difference equations}

In this section we introduce two classes of lattice equations, which will be denoted by ${\cal{F}}$ and ${\cal{F}}^\prime$, respectively. These families contain equations involving the values of a function $u$ at the vertices of an elementary quadrilateral, as shown in Figure \ref{Xenitidis:fig:quad}.
\begin{center}
\begin{figure}[ht]
\begin{minipage}{14pc}
\centertexdraw{ \setunitscale 0.5
\linewd 0.02 \arrowheadtype t:F 
\htext(0 0.5) {\phantom{T}}
\move (-1 -2) \lvec (1 -2) 
\move(-1 -2) \lvec (-1 0) \move(1 -2) \lvec (1 0) \move(-1 0) \lvec(1 0)
\move (1 -2) \fcir f:0.0 r:0.1 \move (-1 -2) \fcir f:0.0 r:0.1
 \move (-1 0) \fcir f:0.0 r:0.1 \move (1 0) \fcir f:0.0 r:0.1  
\htext (-1.1 -2.5) {$u_{(0,0)}$} \htext (.9 -2.5) {$u_{(1,0)}$} \htext (0 -2.25) {$\alpha$}
\htext (-1.1 .15) {$u_{(0,1)}$} \htext (.9 .15) {$u_{(1,1)}$} \htext (0 .1) {$\alpha$}
\htext (-1.25 -1) {$\beta$} \htext (1.1 -1) {$\beta$}}
\caption{} \label{Xenitidis:fig:quad}
\end{minipage}\hspace{1.pc}%
\begin{minipage}{14pc}
\centertexdraw{ \setunitscale 0.5
\linewd 0.02 \arrowheadtype t:F 
\htext(0 0.5) {\phantom{T}}
\move (-1 -2) \lvec (1 -2) 
\move(-1 -2) \lvec (-1 0) \move(1 -2) \lvec (1 0) \move(-1 0) \lvec(1 0)
\move (1 -2) \fcir f:0.0 r:0.1 \move (-1 -2) \fcir f:0.0 r:0.1
 \move (-1 0) \fcir f:0.0 r:0.1 \move (1 0) \fcir f:0.0 r:0.1  
 \move (-1 -2) \lpatt (.1 .1) \lvec (1 0) \move (1 -2) \lvec (-1 0) \lpatt ()
\htext (-1.1 -2.5) {$u_{(0,0)}$} \htext (.9 -2.5) {$u_{(1,0)}$} \htext (0 -2.35) {$h_{34}$}
\htext (-1.1 .15) {$u_{(0,1)}$} \htext (.9 .15) {$u_{(1,1)}$} \htext (0 .1) {$h_{12}$}
\htext (-1.45 -1) {$h_{24}$} \htext (1.1 -1) {$h_{13}$}
\htext (-.5 -1.75) {$h_{23}$} \htext (.15 -1.75 ) {$h_{14}$}}
\caption{} \label{Xenitidis:polyhg}
\end{minipage}
\end{figure}
\end{center}
Specifically, the equations belonging in these classes are autonomous and have the form
\begin{equation}
Q(u_{(0,0)},u_{(1,0)},u_{(0,1)},u_{(1,1)};\alpha,\beta) \,=\, 0\,, \label{Xenitidis:eq:genform}
\end{equation}
where the function $Q$ satisfies two basic requirements: it is affine linear and depends explicitly on the four indicated values of $u$, i.e.
$$\partial_{u_{(i,j)}} Q(u_{(0,0)},u_{(1,0)},u_{(0,1)},u_{(1,1)};\alpha,\beta) \, \ne \, 0$$
and 
$$\partial_{u_{(i,j)}}^2 Q(u_{(0,0)},u_{(1,0)},u_{(0,1)},u_{(1,1)};\alpha,\beta)\, =\, 0\,,$$
where $i$, $j$ = 0, 1.

Further restrictions on the function $Q$ provide the conditions which separate the members of the above classes. Specifically, the members of $\cal{F}$ possess the ${\mathrm{D}}_4$-symmetry, i.e. the function $Q$ satisfies the following requirement:
\begin{eqnarray}
Q(u_{(0,0)},u_{(1,0)},u_{(0,1)},u_{(1,1)};\alpha,\beta) &=&  \epsilon Q(u_{(0,0)},u_{(0,1)},u_{(1,0)},u_{(1,1)};\beta,\alpha) \nonumber\\
&=&  \sigma Q(u_{(1,0)},u_{(0,0)},u_{(1,1)},u_{(0,1)};\alpha,\beta) \,, \nonumber\\
& & \label{Xenitidis:condF}
\end{eqnarray}
where $\epsilon = \pm 1$ and $\sigma = \pm 1$.

On the other hand, equations belonging in ${\cal{F}}^\prime$ possess the symmetries:
\begin{eqnarray}
Q(u_{(0,0)},u_{(1,0)},u_{(0,1)},u_{(1,1)})&=& \tau\,Q(u_{(1,0)},u_{(0,0)},u_{(1,1)},u_{(0,1)}) \nonumber\\
&=& \tau^\prime\,Q(u_{(0,1)},u_{(1,1)},u_{(0,0)},u_{(1,0)})\,, \label{Xenitidis:condFp}
\end{eqnarray}
where $\tau\,=\,\pm\,1$ and $\tau^\prime\,=\,\pm\,1$ \footnote{The dependence of $Q$ on the lattice parameters $\alpha$, $\beta$ is not affected by these symmetries. That is why we omit them from the arguments of $Q$.}.

It is easy to prove that, ${\cal F}^\prime$ actually contains ${\cal F}$. To this end, it is sufficient to prove that, every equation in the class ${\cal F}$ also satisfies requirement (\ref{Xenitidis:condFp}). Indeed, supposing $Q(u,x,y,z;\alpha,\beta)$ satisfies (\ref{Xenitidis:condF}), then using the latter relations one easily finds that
$$Q(u,x,y,z;\alpha,\beta)\,=\,\sigma\,Q(x,u,z,y;\alpha,\beta)\,=\,\sigma\,Q(y,z,u,x;\alpha,\beta)\,,$$
i.e. the function $Q$ also satisfies (\ref{Xenitidis:condFp}) with $\tau = \tau^\prime = \sigma$.

The affine linearity of $Q$ implies that one can define six different polynomials in terms of $Q$ and its derivatives \cite{Xenitidis:ABS,Xenitidis:ABS1,Xenitidis:TTX}, four of them assigned to the edges and the rest to the diagonals of the elementary quadrilateral where the equation is defined, as illustrated in Figure \ref{Xenitidis:polyhg}. 

Specifically, each polynomial $h_{i j}$ is defined by
$$h_{ij}\,=\,h_{j\,i}\,:=\,Q\,Q_{,i j}\,-\,Q_{,i}\,Q_{,j}\,,\quad i\,\ne\,j\,,\quad i,\,j\,=\,1,\ldots,\,4,$$
where $Q_{,i}$ denotes the derivative of $Q$ with respect to its $i$-th argument and $Q_{,i j}$ the second order derivative $Q$ with respect to its $i$-th and $j$-th argument. It is a bi-quadratic polynomial in the values of $u$ assigned to the end-points of the edge or diagonal at which it corresponds. Finally, the relations
\begin{equation}h_{12}\,h_{34}\,=\,h_{13}\,h_{24}\,=\,h_{14}\,h_{23} \label{Xenitidis:relhG} \end{equation}
hold in view of the equation $Q\,=\,0$.

The imposed symmetries of $Q$ yield further restrictions on $h_{i j}$. In particular, taking into account the symmetries on $Q$ for each class separately, one arrives at the following conclusions.
\begin{enumerate}
\item Class of equations ${\cal{F}}$ 
\begin{enumerate}
\item The polynomials on the edges may be given in terms of a polynomial $h(x,y;\alpha,\beta)$, which is symmetric in its two first arguments. Specifically,
\begin{equation}
h_{ij}(x,y)\,=\,\left\{\begin{array}{l c} h(x,y;\alpha,\beta), & |i-j|\,=\,1 \\ h(x,y;\beta,\alpha), & |i-j|\,=\,2 \end{array} \right.\,,
\end{equation}
where $i \ne j$  and $\{i,j\} \ne \{2,3\}$.
\item The diagonal polynomials have the same form
\begin{equation}h_{14}(x,y)\,=\,h_{23}(x,y)\,=\, G(x,y;\alpha,\beta)\,, \end{equation}
where $G$ is symmetric in its two first arguments and in the parameters.
\end{enumerate}
\item Class of equations ${\cal{F}}^\prime$
\begin{enumerate}
\item The polynomials on the edges read as
\begin{equation}
h_{ij}(x,y)\,=\,\left\{\begin{array}{l c} h_1(x,y), & |i-j|\,=\,1 \\ h_2(x,y), & |i-j|\,=\,2 \end{array} \right.\,,
\end{equation}
where $i \ne j$, $\{i,j\} \ne \{2,3\}$. In the above relation, $h_1$, $h_2$ are biquadratic, symmetric polynomials of their arguments.
\item The diagonal polynomials have the same form, i.e.
\begin{equation}h_{14}(x,y)\,=\,h_{23}(x,y)\,=\, G(x,y)\,, \end{equation}
with $G$ being symmetric.
\end{enumerate}
\end{enumerate} 

The most generic equation (\ref{Xenitidis:eq:genform}) belonging in ${\cal F}^\prime$ is equation {\bf Q5}, \cite{Xenitidis:Viallet}, for which the function $Q$ has the following form.
\begin{eqnarray}
 Q(u,x,y,z):= a_1 u x y z  + a_2  \left(u y z + x y z + u x z + u x y\right) + a_3 \left(u x + y z\right) &&\nonumber\\
+ a_4 \left(u z + x y\right) + a_5 \left(x z + u y\right) + a_6\,\left(u + x + y + z\right) + a_7, &&  \label{Xenitidis:Q5}
\end{eqnarray}
where $a_i$ are free parameters.

Equation {\bf Q5} contains equations of the class ${\cal{F}}$ by choosing appropriately the free parameters $a_i$. The specific choices leading to the ABS equations {\bf H1}--{\bf H3} and {\bf Q1}--{\bf Q4} are presented in the following list. \\

\hspace{-.7cm} {\it i}) Equations {\bf H1}--{\bf H3} and {\bf Q1}--{\bf Q3} correspond to the choice $a_1=a_2=0$, while the rest parameters are chosen according to the next table.
\begin{center}
\begin{tabular}{llllll} \hline
 				&  $a_3$ & $a_4$ & $a_5$   & $a_6$                    & $a_7$ \\ \hline
{\bf H1}&  1   &   0   &   $-1$  &   0                      & $\beta-\alpha$ \\ \hline
{\bf H2}&  1   &   0   &   $-1$  &   $\beta-\alpha$         & $\beta^2-\alpha^2$ \\ \hline
{\bf H3}& $\alpha$&   0   &$-\beta$&  0                       & $\delta(\alpha^2-\beta^2)$ \\ \hline
{\bf Q1}& $\alpha$&$\beta-\alpha$&$-\beta$&  0                       & $\delta \alpha \beta (\alpha-\beta)$ \\ \hline
{\bf Q2}& $\alpha$&$\beta-\alpha$&$-\beta$&$\alpha \beta (\alpha-\beta)$& {\tiny{$\alpha \beta (\beta-\alpha) (\alpha^2-\alpha \beta + \beta^2)$}} \\ \hline
{\bf Q3}& {\footnotesize{$\beta (\alpha^2-1)$}}&{\footnotesize{$\beta^2-\alpha^2$}}&{\footnotesize{$\alpha(1-\beta^2)$}}&0& {\footnotesize{$\frac{-\delta (\alpha^2- \beta^2) (\alpha^2-1) (\beta^2-1)}{4 \alpha \beta}$}} \\ \hline
\end{tabular}
\end{center}

\hspace{-.6cm} {\it ii}) Equation {\bf Q4}, which is the master equation of the members of the ABS list, \cite{Xenitidis:AS}, corresponds to the following choices for the parameters:
\begin{eqnarray*}
&& a_1 = a+b \,,\quad a_2=-a \beta - b \alpha\,,\quad a_3 = \frac{a b (a+b)}{2 (\alpha-\beta)} + a \beta^2 - (2 \alpha^2 - \frac{g_2}{4}) b\,,\\
&& a_4=a \beta^2 + b \alpha^2\,,\quad a_5 = \frac{a b (a+b)}{2 (\beta-\alpha)} + b \alpha^2 - (2 \beta^2 - \frac{g_2}{4}) a\,, \\
&& a_6 = \frac{g_3}{2}a_1 - \frac{g_2}{4} a_2\,,\,\,\,a_7=\frac{g_2^2}{16}a_1-g_3 a_2\,,
\end{eqnarray*}
with $a^2\, =\, p(\alpha)$, $b^2 \,=\, p(\beta)$, where $p(x)\,=\,4 x^3-g_2 x - g_3$.

\begin{remark}
Equations {\bf H1}, ${\bf H3}_{\delta =0}$ and ${\bf Q1}_{\delta=0}$ are also referred to as the discrete potential, modified and Schwarzian KdV equation, respectively, cf. review \cite{Xenitidis:NC}. Equations {\bf Q1} and ${\bf Q3}_{\delta=0}$, are related to Nijhoff-Quispel-Capel (NQC) equation, \cite{Xenitidis:NQC}, cf. also \cite{Xenitidis:ABS,Xenitidis:AHN2}. Equation {\bf Q4} was first presented by Adler in \cite{Xenitidis:Adler1}, as the superposition principle of the B{\"a}cklund transformation for the Krichever-Novikov equation.
\end{remark}

\section{Integrability aspects of the ABS equations: auto-B{\"a}cklund transformations and Lax pairs}

\begin{center}
\begin{figure}[h]
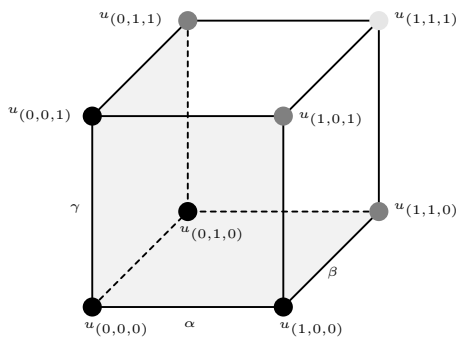

\centertexdraw{ \setunitscale 0.5
\linewd 0.02 \arrowheadtype t:F 
\htext(0 0.5) {\phantom{T}}
\move(0 0) \lvec(0 2) \lvec(1 3) \lvec(1 1) \lvec(0 0) \ifill f:0.95 
\move(0 0) \lvec(2 0) \lvec(3 1) \lvec(1 1) \lvec(0 0) \ifill f:0.95 
\move(0 0) \lvec(2 0) \lvec(2 2) \lvec(0 2) \lvec(0 0) \ifill f:0.95 
\move(0 0) \lvec(2 0) \lvec(2 2) \lvec(0 2) \lvec(0 0)
\move(1 3) \lvec(3 3) \lvec(3 1) \lpatt(.05 .05) \lvec(1 1) \lvec(1 3)
\move(0 0) \lvec(1 1) \lpatt() \move(0 2) \lvec(1 3) \move(2 2) \lvec(3 3) \move(2 0) \lvec(3 1)
\move (0 0) \fcir f:0.0 r:0.1 \move (2 0) \fcir f:0.0 r:0.1 \move (1 1) \fcir f:0.0 r:0.1 \move (0 2) \fcir f:0.0 r:0.1 
\move (2 2) \fcir f:0.5 r:0.1 \move (1 3) \fcir f:0.5 r:0.1 \move (3 1) \fcir f:0.5 r:0.1  
\move(3 3) \fcir f:0.9 r:0.1
\htext (-0.1 -.3) {\tiny{$u_{(0,0,0)}$}} \htext (1.95 -.3) {\tiny{$u_{(1,0,0)}$}} \htext(0.95 -.2) {\tiny{$\alpha$}}
\htext (0.9 .7) {\tiny{$u_{(0,1,0)}$}} \htext(2.45 .3) {\tiny{$\beta$}}
\htext(-.9 1.95) {\tiny{$u_{(0,0,1)}$}} \htext(-.25 1) {\tiny{$\gamma$}}
\htext(3.15 .95) {\tiny{$u_{(1,1,0)}$}} \htext (0.05 2.95) {\tiny{$u_{(0,1,1)}$}} 
\htext(2.15 1.9) {\tiny{$u_{(1,0,1)}$}} 
\htext(3.15 2.95) {\tiny{$u_{(1,1,1)}$}}
}
\caption{Three dimensional consistency} \label{Xenitidis:3Dcons}
\end{figure}
\end{center}

An important characteristic of the ABS equations is their multidimensional consistency \cite{Xenitidis:ABS}. The latter means that, the equation can be consistently embedded on a three dimensional lattice, as illustrated in Figure \ref{Xenitidis:3Dcons}, and, consequently, on a multidimensional lattice. 

To be more precise, suppose that, the function $u$ depends on a third lattice variable $k$, with which the parameter $\gamma$ is associated. Now, each face of the cube carries a copy of the equation, cf. Figure \ref{Xenitidis:3Dcons}, which involves the values of $u$ assigned to the vertices of the face, as well as the corresponding parameters assigned to the edges. Thus, starting with the values of $u$ at the black vertices and using the equations on the gray faces, one can uniquely evaluate the values of $u$ at the gray vertices of the cube. Subsequently, the equations on the white faces provide three different ways to calculate the value $u_{(1,1,1)}$. If these different routes lead to the same result, than the equation is said to possess the consistency property. Moreover, if the resulting value $u_{(1,1,1)}$ is independent of $u_{(0,0,0)}$, then the equation fulfills the tetrahedron property.

The ABS equations possess both of the above properties, \cite{Xenitidis:ABS}. The consistency and the tetrahedron properties imply that, the polynomial $h$ related to the edges is factorized as
$$h(x,y;\alpha,\beta)\,=\,k(\alpha,\beta) \,f(x,y,\alpha)\,,$$ 
where the function $k(\alpha,\beta)$ is antisymmetric, i.e.
$$k(\beta,\alpha) \,= \, - k(\alpha,\beta)\,.$$
Furthermore, the discriminant
$$d(x)\, =\, f_{,y}^2 \,-\, 2\, f\, f_{,y y}$$
is independent of the lattice parameters. The functions $f$, $k$ and $G$ corresponding to each one of the equations under consideration are presented in Appendix A.

The consistency property, which may be thought as the discrete analog of the hierarchy of commuting flows of integrable differential equations, \cite{Xenitidis:NW}, can be used effectively to derive B{\"a}cklund transformations and Lax pairs, \cite{Xenitidis:Nij1,Xenitidis:BobSuris, Xenitidis:Atkinson}. In fact, for the case of the ABS equations, this leads to the following result.

\begin{proposition}
The system
\begin{equation}
{\mathds{B}}_d(u,\tilde{u},\lambda) \,:=\,\left\{ \begin{array}{l} Q(u_{(0,0)},u_{(1,0)},\tilde{u}_{(0,0)},\tilde{u}_{(1,0)};\alpha,\lambda) \,=\, 0\\
Q(u_{(0,0)},u_{(0,1)},\tilde{u}_{(0,0)},\tilde{u}_{(0,1)};\beta,\lambda) \,=\, 0 \end{array} \right. \label{Xenitidis:discautoBac}
\end{equation}
defines an auto-B{\"{a}}cklund transformation for the ABS equation
\begin{equation}Q(u_{(0,0)},u_{(1,0)},u_{(0,1)},u_{(1,1)};\alpha,\beta)\,=\,0\,. \label{Xenitidis:autoBacdiseq} \end{equation}
\end{proposition}

\begin{proof}
Let us first recast equations (\ref{Xenitidis:discautoBac}) and (\ref{Xenitidis:autoBacdiseq}) in terms of the polynomial $f$ by using relations (\ref{Xenitidis:relhG}), i.e. rewrite them as
\begin{eqnarray}
f(u_{(0,0)},u_{(1,0)},\alpha) f(\tilde{u}_{(0,0)},\tilde{u}_{(1,0)},\alpha)= f(u_{(0,0)},\tilde{u}_{(0,0)},\lambda) f(u_{(1,0)},\tilde{u}_{(1,0)},\lambda), \label{Xenitidis:discautoBac1a} \\
f(u_{(0,0)},u_{(0,1)},\beta) f(\tilde{u}_{(0,0)},\tilde{u}_{(0,1)},\beta)= f(u_{(0,0)},\tilde{u}_{(0,0)},\lambda) f(u_{(0,1)},\tilde{u}_{(0,1)},\lambda), \label{Xenitidis:discautoBac1b}
\end{eqnarray}
and
\begin{eqnarray}
f(u_{(0,0)},u_{(1,0)},\alpha) f(u_{(0,1)},u_{(1,1)},\alpha) = f(u_{(0,0)},u_{(0,1)},\beta) f(u_{(1,0)},u_{(1,1)},\beta), \label{Xenitidis:autoBacdiseq11}
\end{eqnarray}
respectively.

Let $u$ be a solution of (\ref{Xenitidis:autoBacdiseq11}). We solve equations (\ref{Xenitidis:discautoBac1a}), (\ref{Xenitidis:discautoBac1b}) for the polynomials $f(u_{(0,0)},u_{(1,0)},\alpha)$ and $f(u_{(0,0)},u_{(0,1)},\beta)$, respectively, and take the shift of the resulting relations in the $m$ and $n$ direction, respectively. The substitution of all these relations into (\ref{Xenitidis:autoBacdiseq11}) leads to
\begin{eqnarray}
f(\tilde{u}_{(0,0)},\tilde{u}_{(1,0)},\alpha) f(\tilde{u}_{(0,1)},\tilde{u}_{(1,1)},\alpha)=f(\tilde{u}_{(0,0)},\tilde{u}_{(0,1)},\beta) f(\tilde{u}_{(1,0)},\tilde{u}_{(1,1)},\beta), \label{Xenitidis:autoBacdiseq12}
\end{eqnarray}
which is equivalent to
$$Q(\tilde{u}_{(0,0)},\tilde{u}_{(1,0)},\tilde{u}_{(0,1)},\tilde{u}_{(1,1)};\alpha,\beta)\,=\,0\,.$$

Conversely, assume that $\tilde{u}$ satisfies the above equation. We solve (\ref{Xenitidis:discautoBac1a}), (\ref{Xenitidis:discautoBac1b}) for the polynomials $f(\tilde{u}_{(0,0)},\tilde{u}_{(1,0)},\alpha)$ and $f(\tilde{u}_{(0,0)},\tilde{u}_{(0,1)},\beta)$, respectively. The resulting expressions and their shifts, combined with (\ref{Xenitidis:autoBacdiseq12}), lead to equation (\ref{Xenitidis:autoBacdiseq11}).
\end{proof}
\begin{figure}[h]
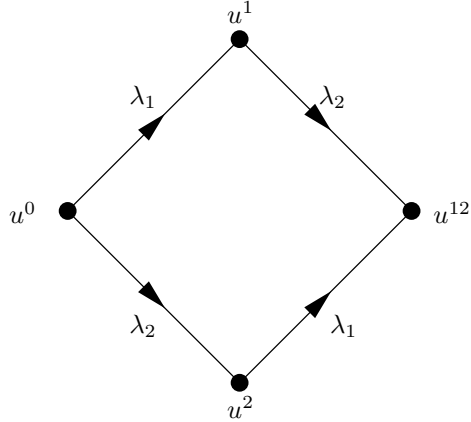

\centertexdraw{ \setunitscale .6 \linewd 0.01 \arrowheadtype t:F
\move(0 0) \lvec (1.5 1.5) \lvec(3 0) \lvec(1.5 -1.5) \lvec(0 0) 
\move(0 0) \fcir f:0.0 r:0.08
\move(1.5 1.5) \fcir f:0.0 r:0.08
\move(3 0) \fcir f:0.0 r:0.08
\move(1.5 -1.5) \fcir f:0.0 r:0.08
\move(.8 .8) \avec(.85 .85)
\move(2.25 .75) \avec(2.3 .7)
\move(.8 -.8) \avec(.85 -.85)
\move(2.25 -.75) \avec(2.3 -.7) 
\htext (-0.5 -0.1) {$u^{0}$} \htext (1.4 1.65) {$u^1$} \htext (3.2 -.1) {$u^{1 2}$} \htext (1.4 -1.8) {$u^2$}
\htext(0.55 0.9){$\lambda_1$} \htext(0.55 -1.1){$\lambda_2$} \htext(2.2 0.9){$\lambda_2$} \htext(2.3 -1.1){$\lambda_1$}}
\caption{Bianchi commuting diagramm} \label{Xenitidis:Bianchicomd}
\end{figure}

A direct consequence of the previous result is the next proposition.
\begin{proposition}[Bianchi commuting diagramm] \label{Xenitidis:Bianchidiscrete}
Let $u^1$, $u^2$ be solutions of the ABS equation (\ref{Xenitidis:autoBacdiseq}), generated by means of the auto-B{\"a}cklund tranformation ${\mathds{B}}_d$ from a known solution $u^0$ via the B{\"a}cklund parameters $\lambda_1$ and $\lambda_2$, respectively. Then, there is a third solution $u^{12}$, which is given algebraically by
$$Q\left(u^0,u^1,u^2,u^{12};\lambda_1,\lambda_2\right)\,=\,0\,,$$
and is constructed according to Figure \ref{Xenitidis:Bianchicomd}. 
\end{proposition}

\begin{proof}
Rewriting the equations of the auto-B{\"a}cklund transformation in terms of the polynomial $f$, we divide the equations of ${\mathds{B}}_d\left(u^0,u^1,\lambda_1\right)$ by the corresponding ones of ${\mathds{B}}_d\left(u^0,u^2,\lambda_2\right)$ and, the equations of ${\mathds{B}}_d\left(u^1,u^{12},\lambda_2\right)$ by the corresponding ones of ${\mathds{B}}_d\left(u^2,u^{12},\lambda_1\right)$. Equating the resulting relations, we get
\begin{equation}F\,{\cal{S}}_n(F)\,=\,G\,{\cal{S}}_n(G)\,,\qquad F\,{\cal{S}}_m(F)\,=\,G\,{\cal{S}}_m(G)\,,\label{Xenitidis:cond1Bia}\end{equation}
where
\begin{eqnarray*}
F &=& f(u^0_{(0,0)},u^1_{(0,0)},\lambda_1) f(u^2_{(0,0)},u^{12}_{(0,0)},\lambda_1)\,,\\
G &=& f(u^0_{(0,0)},u^2_{(0,0)},\lambda_2) f(u^1_{(0,0)},u^{12}_{(0,0)},\lambda_2)\,.
\end{eqnarray*}

From equations (\ref{Xenitidis:cond1Bia}) we get that $F=G$. The latter equivalently implies that
$$Q\left(u^0,u^1,u^2,u^{12};\lambda_1,\lambda_2\right)\,=\,0\,.$$
\end{proof}

\begin{remark}
Auto-B{\"a}cklund transformations different from the ones presented here, as well as some hetero-B{\"a}cklund transformations, were given recently by Atkinson in \cite{Xenitidis:Atkinson}.
\end{remark}

The consistency property also serves in the construction of a Lax pair, through an algorithmic procedure \cite{Xenitidis:Nij1,Xenitidis:BobSuris,Xenitidis:LeviYam}. Alternatively, using the fact that, a B{\"a}cklund transformation may be regarded as a gauge transformation for the Lax pair, \cite{Xenitidis:Crampin}, one may construct a Lax pair for the ABS equations using the equations constituting ${\mathds{B}}_d$ \cite{Xenitidis:Xen}. The two approaches are equivalent and lead essentially to the same result, which can be stated in the form of the following proposition.

\begin{proposition}
The equation of the ABS class
$$Q(u_{(0,0)},u_{(1,0)},u_{(0,1)},u_{(1,1)};\alpha,\beta)\,=\,0$$
arises in the compatibility condition of the linear system
\begin{equation}
\Psi_{(1,0)}={\rm{L}}(u_{(0,0)},u_{(1,0)};\alpha,\lambda) \Psi_{(0,0)},\,\,\Psi_{(0,1)}={\rm{L}}(u_{(0,0)},u_{(0,1)};\beta,\lambda) \Psi_{(0,0)},
\end{equation}
where
\begin{equation}
{\rm{L}}(x^1,x^2;a,\lambda)\,=\,\frac{1}{\sqrt{k(a,\lambda)\,f(x^1,x^2,a)}}\,\left( \begin{array}{c c} Q_{,4} & -Q_{,3 4}\\Q & -Q_{,3} \end{array} \right)\,, \label{Xenitidis:Lmatrix}
\end{equation}
and $Q\,=\,Q(x^1,x^2,x^3,x^4;a,\lambda)$ and its derivatives ($Q_{,i}\,=\,\partial_{x^i}Q$) are evaluated at $x^3=x^4=0$.
\end{proposition}

\section{Infinite hierarchies of generalized symmetries}

In this section we present the symmetry analysis of the equations belonging in the class ${\cal{F}}^\prime$, while the corresponding analysis for the class $\cal{F}$ was given in \cite{Xenitidis:TTX}. The conclusions of the symmetry analysis are summarized in the form of two propositions and, subsequently, are applied to equation {\bf Q5}. Finally, we explicitly construct infinite hierarchies of generalized symmetries for all of the ABS equations, the members of which are constructed inductively using linear differential operators.

Let us first present the two propositions about the symmetries of the equations belonging in ${\cal{F}}^\prime$.
\begin{proposition}
Every equation in the class ${\cal F}^\prime$ admits two three point generalized symmetries with generators the vector fields
$${\mathbf{v}}_n = \left( \frac{h_1(u_{(0,0)},u_{(1,0)})}{u_{(1,0)}-u_{(-1,0)}} - \frac{1}{2} 
{h_1}_{,u_{(1,0)}}(u_{(0,0)},u_{(1,0)}) \right) \partial_{u_{(0,0)}}\,, $$
and
$${\mathbf{v}}_m = \left( \frac{h_2(u_{(0,0)},u_{(0,1)})}{u_{(0,1)}-u_{(0,-1)}} - \frac{1}{2} 
{h_2}_{,u_{(0,1)}}(u_{(0,0)},u_{(0,1)}) \right) \partial_{u_{(0,0)}}\,,$$
respectively.
\end{proposition}

\begin{proposition}
Let an equation in the class ${\cal F}^\prime$ be such that, the matrices
\begin{equation}
{\mathcal{G}}_i\,=\,\left. \left(\begin{array}{ccc}
h_i(x,y) & G(x,z) & G(x,w) \\
{h_i}_{,x}(x,y) & G_{,x}(x,z) & G_{,x}(x,w) \\
{h_i}_{,x x}(x,y) & G_{,x x}(x,z) & G_{,x x}(x,w)\end{array}
 \right)\right|_{x\,=\,0}\,,\quad i\,=\,1,\,2\,, \label{Xenitidis:eq:matrixGi}
\end{equation}
are invertible. Then , the generator of any five point generalized symmetry of this equation will be necessarily a vector field of the form
$$ {\bf v} \,=\, a(n) {\mathbf{v}}_n + b(m) {\mathbf{v}}_m + \frac{1}{2}\,\psi(n,m,u_{(0,0)}) \partial_{u_{(0,0)}}\,,$$
where the functions $a(n)$, $b(m)$ and $\psi(n,m,u_{(0,0)})$ satisfy the determining equation
\begin{eqnarray*}
\left( a(n) - a(n+1) \right) \,h_1(u_{(0,0)},u_{(1,0)})^2\,\partial_{u_{(1,0)}} \left( \frac{G(u_{(1,0)},u_{(0,1)})}{h_1(u_{(0,0)},u_{(1,0)})} \right)
&& \\
+ \, \left( b(m) - b(m+1) \right) \,h_2(u_{(0,0)},u_{(0,1)})^2 \,\partial_{u_{(0,1)}} \left( \frac{G(u_{(1,0)},u_{(0,1)})}{h_2(u_{(0,0)},u_{(0,1)})} \right) && \\
&&\\
+ G(u_{(1,0)},u_{(0,1)}) \psi(n,m,u_{(0,0)})  + h_2(u_{(0,0)},u_{(0,1)}) \psi(n+1,m,u_{(1,0)}) && \nonumber\\
&& \nonumber \\
+ h_1(u_{(0,0)},u_{(1,0)}) \psi(n,m+1,u_{(0,1)}) - Q_{,u_{(1,1)}}^2 \psi(n+1,m+1,u_{(1,1)})  = 0 \,.
\nonumber \end{eqnarray*}
\end{proposition} 

\begin{proof}
The proofs of the above propositions follow from the corresponding ones given in \cite{Xenitidis:TTX} by making the following changes
\begin{eqnarray*}
h(u,x,\alpha,\beta) \longrightarrow h_1(u,x)\,,\quad h(u,x,\beta,\alpha) \longrightarrow h_2(u,x)\,.
\end{eqnarray*}
\end{proof}

The application of the above symmetry analysis to {\bf Q5} implies that, it admits only a pair of three point generalized symmetries with generators the vector fields
\begin{eqnarray*}
{\mathbf{v}}_n &=& \frac{h_1(u_{(0,0)},u_{(1,0)})}{u_{(1,0)}-u_{(-1,0)}}- \frac{1}{2} \frac{\partial h_1(u_{(0,0)},u_{(1,0)})}{\partial u_{(1,0)}}\,,\\ {\mathbf{v}}_m  &=& \frac{h_2(u_{(0,0)},u_{(0,1)})}{u_{(0,1)}-u_{(0,-1)}}-\frac{1}{2} \frac{\partial h_2(u_{(0,0)},u_{(0,1)})}{\partial u_{(0,1)}}\,, \end{eqnarray*}
where
\begin{eqnarray*} 
h_1(x,y) &= & (a_3+a_1 x y+a_2 (x+y)) (a_7+a_3 x y+a_6 (x+y))\,- \nonumber \\
          & &  (a_6+a_5 x+(a_4+a_2 x) y) (a_6+a_4 x+(a_5+a_2 x) y) \,,\\
& &\nonumber \\
h_2(x,y) &=& (a_5+a_1 x y+a_2 (x+y)) (a_7+a_5 x y+a_6 (x+y))\,- \nonumber \\
& & (a_6+a_4 x+(a_3+a_2 x) y)(a_6+a_3 x+(a_4+a_2 x) y)\,.
\end{eqnarray*}

Furthermore, it can be shown by a direct calculation that, the commutator of these two symmetry generators is a trivial symmetry, \cite{Xenitidis:Olver}, i.e. the characteristic of the resulting vector field vanishes on solutions of equation {\bf Q5}. Thus, we can write
\begin{equation}\left[ {\mathbf{v}}_n\,,\,{\mathbf{v}}_m \right]\,=\,0\,. \label{Xenitidis:comrelQ5}\end{equation}
Since {\bf Q5} contains all the ABS equations, the above commutation relation also holds for the corresponding symmetry generators of the latter equations.

On the other hand, the symmetry analysis of the ABS equations, \cite{Xenitidis:TTX}, showed that, all of them admit a pair of three point generalized symmetries, as well as a pair of extended generalized symmetries, the generators of which are the vector fields
\begin{equation}
{\bf v}_n\,\equiv \,{\bf v}_n^{[0]}\,=\,{\rm R}^{[0]}_n\,\partial_{u_{(0,0)}} \,, \quad {\bf v}_m\,\equiv \,{\bf v}_m^{[0]}\,=\,{\rm R}^{[0]}_m\,\partial_{u_{(0,0)}}\,, \label{Xenitidis:genvnvm1}
\end{equation}
and 
\begin{equation}\label{Xenitidis:extgenv1v21}
{\mathbf{V}}_n \,=\, n\,{\rm R}^{[0]}_n\,\partial_{u_{(0,0)}} \,-\, r(\alpha)\, \partial_\alpha\,,\quad {\mathbf{V}}_m \,=\,  m\,{\rm R}^{[0]}_m\,\partial_{u_{(0,0)}}\,-\, r(\beta)\, \partial_\beta\,,
\end{equation}
respectively, with
\begin{eqnarray*}
{\rm R}^{[0]}_n &=& \frac{f(u_{(0,0)},u_{(1,0)},\alpha)}{u_{(1,0)}-u_{(-1,0)}} - \frac{1}{2} f_{,u_{(1,0)}}(u_{(0,0)},u_{(1,0)},\alpha)\,,\\
{\rm R}^{[0]}_m &=& \frac{f(u_{(0,0)},u_{(0,1)},\beta)}{u_{(0,1)}-u_{(0,-1)}} - \frac{1}{2} f_{,u_{(0,1)}}(u_{(0,0)},u_{(0,1)},\beta)\,.
\end{eqnarray*}
The form of the function $r$ appearing in (\ref{Xenitidis:extgenv1v21}) depends on the particular equation and is given in the following table. 
\begin{center}
\begin{tabular}{c c c c c c c c}
\hline Equation & {\bf H1} & {\bf H2} & {\bf H3} & {\bf Q1} & {\bf Q2} & {\bf Q3} & {\bf Q4}  \\
 $r(x)$ & $1$ & $1$ & $-\,\frac{x}{2}$ & $1$ & $1$ &  $-\,\frac{x}{2}$ & $-\,\frac{1}{2}\,\sqrt{4 x^3-g_2 x-g_3}$ \\ \hline
\end{tabular}
\end{center}

The generalized symmetries have been used effectually to derive reductions of partial difference equations to discrete analogues of the Painlev{\'e} equations. The first result in this direction was presented by Nijhoff and Papageorgiou in \cite{Xenitidis:Pap2}, where the reduction of {\bf H1} to a discrete analogue of Painlev{\'e} II was given.

The extended symmetries, which act, not only on the dependent variable $u$ but, on the lattice parameters as well, are very important and can be used effectively in two different ways. The first way is in the construction of solutions through symmetry reductions, \cite{Xenitidis:Xen,Xenitidis:TX}. Assuming that the function $u$ depends continuously on the lattice parameters $\alpha$, $\beta$, one may derive solutions of the given difference equation which remain invariant under the action of both of the extended symmetries generated by the vector fields ${\bf V}_n$, ${\bf V}_m$. In this fashion, one is led to a system of differential-difference equations, which can be equivalently written as an integrable system of differential equations. On the one hand, this differential system is related to the so called generating partial differential equations, introduced by Nijhoff, Hone and Joshi in \cite{Xenitidis:NHJ}, cf. also \cite{Xenitidis:TTX1,Xenitidis:TTX2}. On the other hand, some of its similarity solutions reveal new connections between discrete and continuous Painlev{\'e} equations, \cite{Xenitidis:TX}.

The other way to use the extended symmetries was suggested by Rasin and Hydon in \cite{Xenitidis:RHsym}. Specifically, it was pointed out that, the above extended generalized symmetries can be regarded as master symmetries of the corresponding generalized ones. To prove this, one has to show that the following commutation relations hold
$$
\left[{\bf{V}}_n,{\bf{v}}_n^{[0]}\right]\,\ne\,0\,,\quad \left[\left[{\bf{V}}_n,{\bf{v}}_n^{[0]}\right],{\bf{v}}_n^{[0]}\right]\,=\,0\,,
$$
as well as similar relations for the generators ${\bf V}_m$, ${\bf v}_m^{[0]}$. It should be noted that, the commutators $[{\bf V}_n,{\bf v}_m^{[0]}]$, $[{\bf V}_m,{\bf v}_n^{[0]}]$ lead to trivial symmetries because of the relation (\ref{Xenitidis:comrelQ5}), e.g. $[{\bf V}_n,{\bf v}_m^{[0]}]=n [{\bf v}_n^{[0]},{\bf v}_m^{[0]}] $.

Now, writing out explicitly the commutators $[{\bf V}_n,{\bf v}_n^{[0]}]$ and $[{\bf V}_m,{\bf v}_m^{[0]}]$, one arrives at
\begin{eqnarray}
{\bf v}_n^{[1]} &:=& \left[{\bf V}_n,{\bf v}_n^{[0]}\right] \,=\,{\rm R}_n^{[1]}\,\partial_{u_{(0,0)}}\,, \label{Xenitidis:5pg-com1}\\
{\bf v}_m^{[1]} &:=& \left[{\bf V}_m,{\bf v}_m^{[0]}\right] \,=\,{\rm R}_m^{[1]}\,\partial_{u_{(0,0)}}\,, \label{Xenitidis:5pg-com2}
\end{eqnarray}
where
\begin{eqnarray}
{\rm R}_n^{[1]} = \left(\left({\cal{S}}_n {\rm R}_n^{[0]}\right) \partial_{u_{(1,0)}} - \left({\cal{S}}_n^{(-1)} {\rm R}_n^{[0]}\right) \partial_{u_{(-1,0)}}-r(\alpha) \partial_{\alpha}\right) {\rm R}_n^{[0]}\nonumber \\
\nonumber \\
=\,\,\,\, \frac{f(u_{(0,0)},u_{(1,0)},\alpha) f(u_{(0,0)},u_{(-1,0)},\alpha) (u_{(2,0)}-u_{(-2,0)})}{(u_{(1,0)}-u_{(-1,0)})^2 (u_{(2,0)}-u_{(0,0)}) (u_{(-2,0)}-u_{(0,0)})}, \label{Xenitidis:char5pgs-com1}\\
\nonumber \\
{\rm R}_m^{[1]} =  \left( \left({\cal{S}}_m {\rm R}_m^{[0]}\right) \partial_{u_{(0,1)}} - \left({\cal{S}}_m^{(-1)} {\rm R}_m^{[0]}\right) \partial_{u_{(0,-1)}}-r(\beta) \partial_{\beta}\right) {\rm R}_m^{[0]}\nonumber \\
\nonumber \\
=\,\,\,\, \frac{f(u_{(0,0)},u_{(0,1)},\beta) f(u_{(0,0)},u_{(0,-1)},\beta) (u_{(0,2)}-u_{(0,-2)})}{(u_{(0,1)}-u_{(0,-1)})^2 (u_{(0,2)}-u_{(0,0)}) (u_{(0,-2)}-u_{(0,0)})}. \label{Xenitidis:char5pgs-com2}
\end{eqnarray}

Using definitions (\ref{Xenitidis:5pg-com1}-\ref{Xenitidis:5pg-com2}) and the Jacobi identity, it can be verified straightforwardly that:
$$\left[{\bf v}_n^{[1]}\,,\,{\bf v}_m^{[0]} \right] = \left[{\bf v}_n^{[1]}\,,\,{\bf V}_m \right] = \left[{\bf v}_m^{[1]}\,,\,{\bf v}_n^{[0]} \right] = \left[{\bf v}_m^{[1]}\,,\,{\bf V}_n \right] = \left[{\bf v}_n^{[1]}\,,\,{\bf v}_m^{[1]} \right] = 0 .$$
On the other hand, the commutators
$$\left[{\bf v}_n^{[1]}\,,\,{\bf v}_n^{[0]} \right]\,=\,\left[{\bf v}_m^{[1]}\,,\,{\bf v}_m^{[0]} \right]\,=\,0$$
can be verified by using (\ref{Xenitidis:genvnvm1}), (\ref{Xenitidis:5pg-com1}-\ref{Xenitidis:char5pgs-com2}) and the properties of polynomial $f$, i.e. it is biquadratic and symmetric.

The above analysis implies that, the symmetry generators ${\bf V}_n$, ${\bf V}_m$ are master symmetries of ${\bf v}_n^{[0]}$ and ${\bf v}_m^{[0]}$, respectively. Consequently, an infinite hierarchy of generalized symmetries can be constructed in this fashion, the members of which are defined inductively:
\begin{equation}{\bf v}_i^{[k+1]}={\rm R}_i^{[k+1]}\,\partial_{u_{(0,0)}} := \left[ {\bf V}_i,{\bf v}_i^{[k]}\right] ,\quad k=0,1,\ldots \quad {\mbox{and}} \quad i=n,\,m\,. \label{Xenitidis:infhier}\end{equation}
The characteristic ${\rm R}_i^{[k]}$ involves the values of $u$ at $(2 k + 3)$ points in the $i$ direction of the lattice and is determined by applying successively the linear differential operators 
\begin{eqnarray}
{\cal{R}}_n &=& \sum_{\ell=-\infty}^\infty \ell \, \left({\cal{S}}_n^{(\ell)} {\rm R}_n^{[0]}\right) \partial_{u_{(\ell,0)}} \,-\,r(\alpha)\,\partial_\alpha\,, \\
{\cal{R}}_m  &=& \sum_{\ell=-\infty}^\infty \ell\,\left({\cal{S}}_m^{(\ell)} {\rm R}_m^{[0]}\right) \partial_{u_{(0,\ell)}} \,-\,r(\beta)\,\partial_\beta
 \label{Xenitidis:recoper} \end{eqnarray}
on ${\rm R}_i^{[0]}$, i.e.
\begin{equation}{\rm R}_i^{[k]}\,=\,{\cal{R}}_i^{\,k} {\rm R}_i^{[0]}\,,\quad k\,=\,0,\,1\,\ldots\quad{\mbox{and}}\quad i\,=\,n,\,m\,. \label{Xenitidis:infhierrec3} \end{equation}

\begin{remark}
The linear operators ${\cal{R}}_n$, ${\cal{R}}_m$ may be regarded as recursion operators for equations {\bf Q3} and {\bf Q4}. This follows from the fact, \cite{Xenitidis:TTX}, that, these equations admit only the symmetries generated by the vector fields given in (\ref{Xenitidis:genvnvm1}).
\end{remark}

\begin{remark}
It is worth mentioning some previous results on higher order generalized symmetries for the ABS equations. Specifically, a procedure for the construction of hierarchies of generalized symmetries was presented in \cite{Xenitidis:LeviYam}, where the relation of the ABS equations to Yamilov's discrete Krichever--Novikov equation (YdKN), \cite{Xenitidis:Yami1,Xenitidis:yami}, was explored. On the other hand, particular results about {\bf H1} and {\bf Q1}${}_{\delta=0}$ were presented in \cite{Xenitidis:levi-petr} and \cite{Xenitidis:levi-petr-sc}, respectively, the derivation of which based on the associated spectral problem. The advantage of our approach compared to the above outcomes is that, it leads straightforwardly to explicit expressions for the higher order generalized symmetries for all of the ABS equations.
\end{remark}

The correspondence of the generators ${\bf v}_i^{[k]}$ to group of transformations, \cite{Xenitidis:Olver}, leads to hierarchies of integrable differential-difference equations, the first members of which are special cases of YdKN equation, \cite{Xenitidis:LeviYam}, cf. also \cite{Xenitidis:AS}.

Dropping the one of the two indices corresponding to shifts with respect to $n$ or $m$, and denoting by $a$ the corresponding lattice parameter, these hierarchies have the form
\begin{equation}
\frac{{\rm d} u}{{\rm d} \epsilon_k}\,= \,{\cal{R}}^k\left(\frac{f(u,u_1,a)}{u_1-u_{-1}}\,-\,\frac{1}{2}\,f_{,u_1}(u,u_1,a) \right)\,, \quad k \,=\, 0, \,1,\, \ldots, \label{Xenitidis:hierYdKN}
\end{equation}
where $\epsilon_k$ is the corresponding group parameter,
\begin{equation}
{\cal{R}} = \sum_{\ell=-\infty}^\infty \ell \, \left({\cal{S}}^{(\ell)} \left(\frac{f(u,u_1,a)}{u_1-u_{-1}}\,-\,\frac{1}{2}\,f_{,u_1}(u,u_1,a) \right)\right) \partial_{u_\ell} \,-\,r(a)\,\partial_a\,
\end{equation}
and $\cal{S}$ is the shift operator in the corresponding direction.

The corresponding ABS equation, written now in the form
$$
Q(u,u_1,\tilde{u},\tilde{u}_1;a,\lambda)\,=\,0\,,
$$
is an auto-B{\"a}cklund transformation of equations (\ref{Xenitidis:hierYdKN}), \cite{Xenitidis:LeviYam}, while the latter equations admit the following Lax pair
\begin{equation}
\Psi_1 \,=\, {\rm L}(u,u_1;a,\lambda)\,\Psi \,,\quad \frac{{\rm d} \Psi}{{\rm d} \epsilon_k} \,=\, {\rm N}_k \Psi \,, \quad k \,=\, 0, \,1,\, \ldots\,.
\end{equation}
In the above relations, the matrix ${\rm L}(u,u_1;a,\lambda)$ is given by (\ref{Xenitidis:Lmatrix}) and
\begin{equation} 
{\rm N}_k\,:=\,{\cal{R}}^k \left( \frac{1}{u_{1}-u_{-1}}\,{\rm{X}}\,-\,\frac{1}{2}\,\,{\rm{X}}_{,u_1} \right) \,,\quad k \,=\, 0, \,1,\, \ldots\, ,\end{equation}
where
\begin{equation}
{\rm{X}}\,:=\,-\,f(u,u_{1},a)\,\left({\rm L}(u,u_1;a,\lambda)\right)^{-1}\,\partial_u{\rm L}(u,u_1;a,\lambda)\,.
\end{equation}

\section{Conclusions and prespectives}

We have presented some recent results about difference equations, their integrability aspects and their underlying symmetry structure. We have used as a representative example the equations of the ABS classification \cite{Xenitidis:ABS}. The multidimensional consistency property possessed by the ABS equations provides the means to investigate in a systematic way B{\"a}cklund transformations and Lax pairs for all of them. In the same fashion, integrability aspects of consistent systems of difference equations have been derived, e.g. the discrete Boussinesq system, \cite{Xenitidis:TN1,Xenitidis:TN2}, and the elliptic generalization of the discrete KdV equation, \cite{Xenitidis:NPut}.

The extended generalized symmetries admitted by the ABS equations are proven to provide us an effective tool. They play the key role of master symmetries leading to the explicit construction of infinite hierarchies of generalized symmetries. Moreover, the same symmetries can be used for the construction of similarity solutions, which were called continuously invariant solutions in \cite{Xenitidis:TX}, revealing a new link among difference and differential equations. Moreover, continuously invariant solutions of special type have exposed a novel connection among discrete and continuous Painlev{\'e} equations. It would be interesting to study other systems of difference equations from the point of view of extended generalized symmetries and continuously invariant solutions, such as the discrete Boussinesq system, for which their connection to integrable differential equations was established by different means in \cite{Xenitidis:TN1,Xenitidis:TN2}.

\appendix
\section{The characteristic polynomials of the ABS equations}

\begin{enumerate}[{\bf H}\bf 1]
\item  $f(u,x,\alpha) = 1$\\ $G(x,y) = (x-y)^2$\\ $k(\alpha,\beta) = \beta -\alpha$ 
\item $f(u,x,\alpha) = 2(u + x + \alpha)$ \\ $G(x,y) = (x-y)^2 - (\alpha-\beta)^2$ \\ $k(\alpha,\beta) = \beta -\alpha$
\item $f(u,x,\alpha) = u x + \alpha \delta$ \\ $G(x,y) = (y \alpha - x \beta) (y \beta - x \alpha)$ \\ $k(\alpha,\beta) = \alpha^2-\beta^2$
\end{enumerate}

\begin{enumerate}[{\bf Q}\bf 1]
\item  $ f(u,x,\alpha) \,= \,((u-x)^2 - \alpha^2 \delta^2)/\alpha$ \\ $G(x,y) = \alpha \beta \left((x-y)^2 - (\alpha -\beta)^2 \delta^2\right)$ \\
$k(\alpha,\beta) \,=\, -\alpha \beta (\alpha-\beta)$
\item $ f(u,x,\alpha) = ((u-x)^2 - 2 \alpha^2 (u+x) + \alpha^4)/\alpha$ \\
      $ G(x,y) = \alpha \beta \left((x-y)^2 - 2 (\alpha -\beta)^2 (x+y) + (\alpha-\beta)^4\right)$\\
      $k(\alpha,\beta) = -\alpha \beta (\alpha-\beta)$
\item $f(u,x,\alpha) =\frac{-1}{4\alpha(\alpha^2-1)} (4 \alpha (x-\alpha u) (\alpha x - u) + (\alpha^2-1)^2 \delta^2)$ \\
      $G(x,y) = \frac{(\alpha^2-1) (\beta^2-1)}{4 \alpha \beta} \left(4 \alpha \beta (\alpha y-\beta x) (\beta y-\alpha x)+(\alpha^2-\beta^2) \delta^2\right)$\\
      $k(\alpha,\beta) = (\alpha^2 -  \beta^2) (\alpha^2-1) (\beta^2 - 1)$
\item $f(u,x,\alpha) = \left((u x+ \alpha (u+x) + g_2/4)^2 - (u+x+\alpha)(4 \alpha u x-g_3) \right)/a $\\ \\
			$G(x,y) = (a_1 x y + a_2 (x+y) +a_4) (a_4 x y + a_6 (x+y)+a_7)$ \\
      ${\phantom{G(x,y) = }} - (a_2 x y + a_3 y + a_5 x + a_6) (a_2 x y + a_3 x + a_5 y + a_6)$ \\ \\
      $ k(\alpha,\beta) = \,\frac{a b\left(a^2 b + a b^2 +  \left[12 \alpha \beta^2 - g_2 (\alpha+2\beta) - 3 g_3\right] a + \left[12 \beta \alpha^2 - g_2 (\beta + 2 \alpha) - 3 g_3\right] b \right)}{4 (\alpha-\beta)}$
\end{enumerate}

\end{document}